\documentclass[11pt,twoside]{article} 
\usepackage{asp2004}
\usepackage{epsf}
\usepackage{psfig}
\usepackage{lscape} 
\usepackage{float}

\begin{document}
\title{Spectral Analyses of DO White Dwarfs and PG1159 Stars from 
  the Sloan Digital Sky Survey}
\author{S. D. H\"ugelmeyer,$^1$ S. Dreizler,$^1$ K. Werner,$^2$
  A. Nitta,$^3$ S. J. Kleinman,$^3$ and J. Krzesi\'nski$^3$ } 
\affil{$^1$Institut f\"ur Astrophysik, Universit\"at$\,$G\"ottingen,
Friedrich-Hund-Platz~1, D-37077 G\"ottingen, Germany \\
  $^2$Institut f\"ur Astronomie und Astrophysik,
  Universit\"at$\,$T\"ubingen, Sand~1, D-72076 T\"ubingen, Germany\\
  $^3$New Mexico State University, Apache Point Observatory, 2001
  Apache Point Road, P.O. Box 59, Sunspot, NM 88349, USA\\}

\begin{abstract} 
  SDSS (DR1 and DR2) has recently proposed 7 new DO white dwarfs
  as well as 6 new PG1159
  stars. This is a significant increase in the known number of DOs
  and PG1159 stars. Our spectral analyses provide stellar 
  parameters which can then be used to derive constraints for the
  evolution of H-deficient white dwarfs. A comprehensive
  understanding of these objects is still severely hampered 
  by low-number statistics.
\end{abstract}

\section{Introduction}
DO white dwarfs and PG1159 stars are situated on the upper part of the
white dwarf cooling sequence. Both types of stars are helium-rich and
therefore seem to have undergone an untypical evolutionary path. DO
white dwarfs have effective temperatures from T$_{{\rm eff}}
\approx 45\,000$ K up to T$_{{\rm eff}} \approx 120\,000$ K. PG1159
stars feature temperatures which exceed T$_{{\rm eff}}=65\,000 $ K with an
upper limit of T$_{{\rm eff}}=200\,000 $~K. All these attributes in turn
suggest an evolutionary connection in which PG1159 stars are the
proposed precursors of DO white dwarfs.

\section{Spectral Analyses of DO White Dwarfs and PG1159 Stars}

\begin{figure}[H]
  \plotfiddle{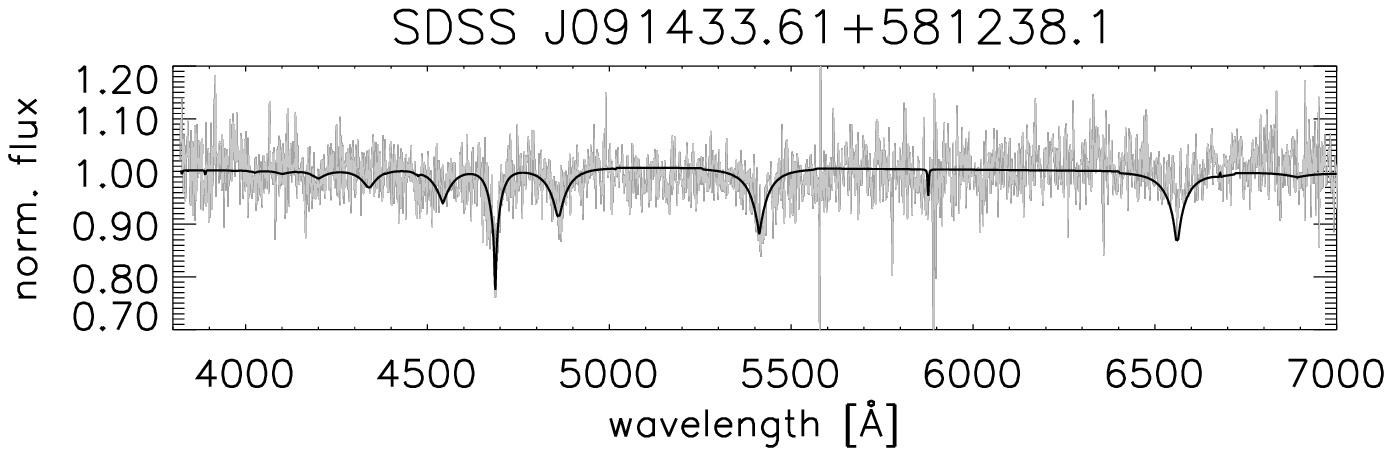}{5cm}{0}{47}{47}{-223}{-90}
  \plotfiddle{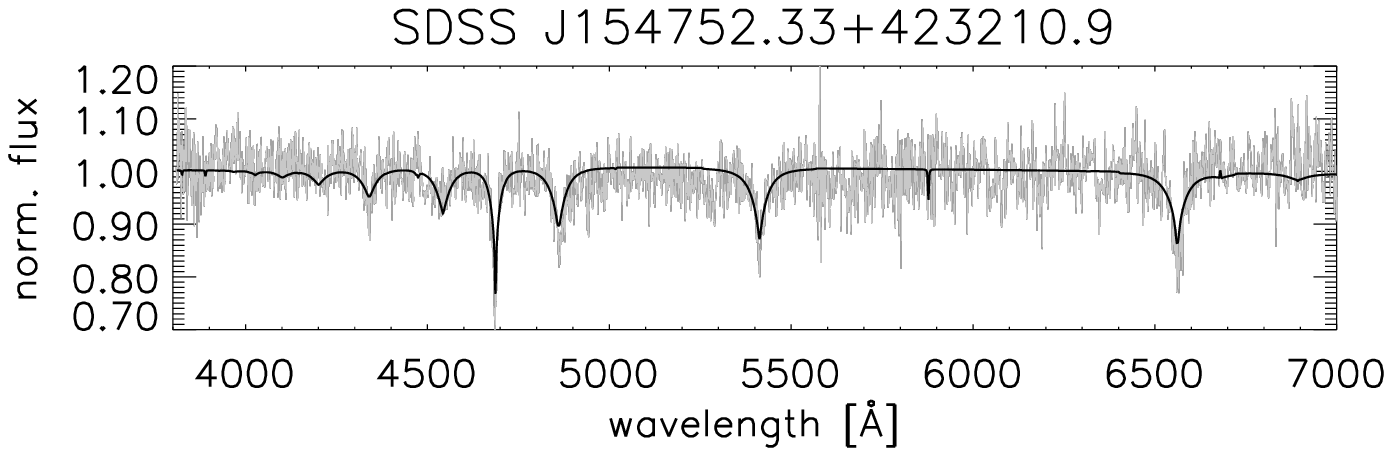}{5cm}{0}{47}{47}{-38}{65}
  \plotfiddle{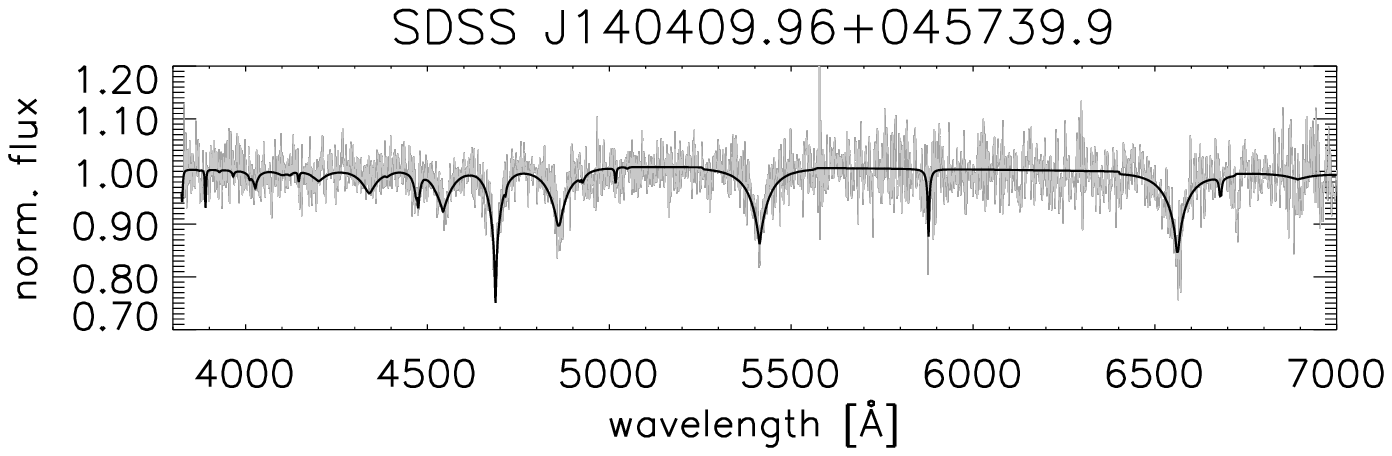}{5cm}{0}{47}{47}{-223}{150}
  \plotfiddle{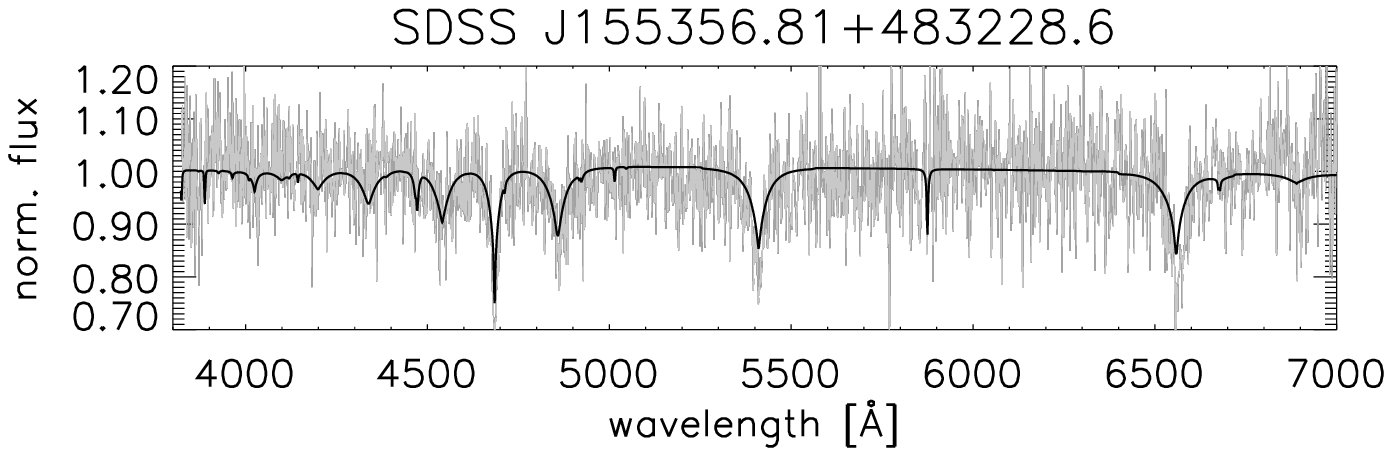}{5cm}{0}{47}{47}{-38}{305}
  \plotfiddle{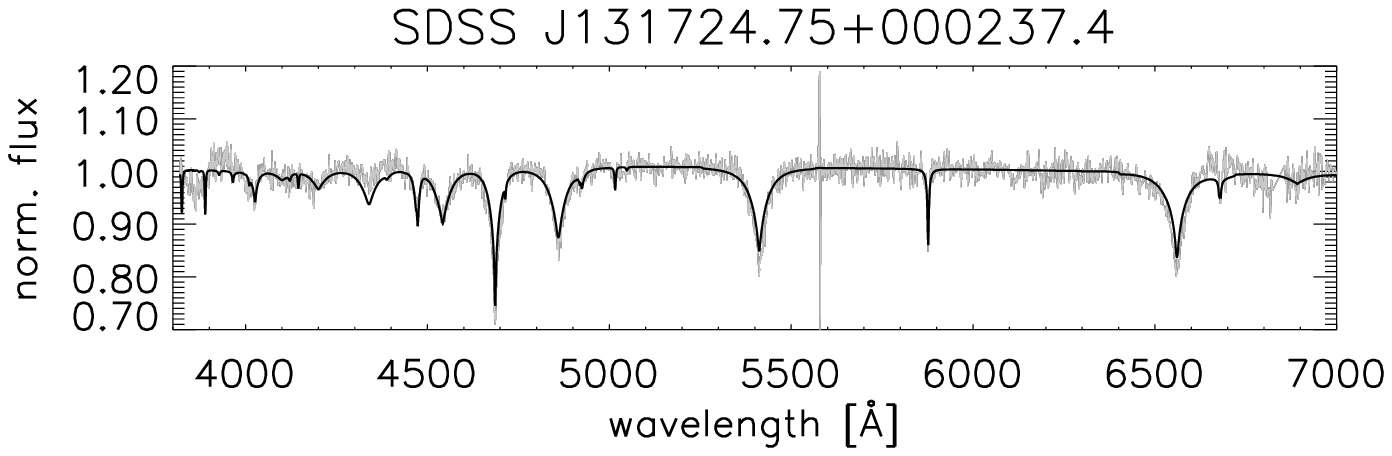}{5cm}{0}{47}{47}{-223}{395}
  \plotfiddle{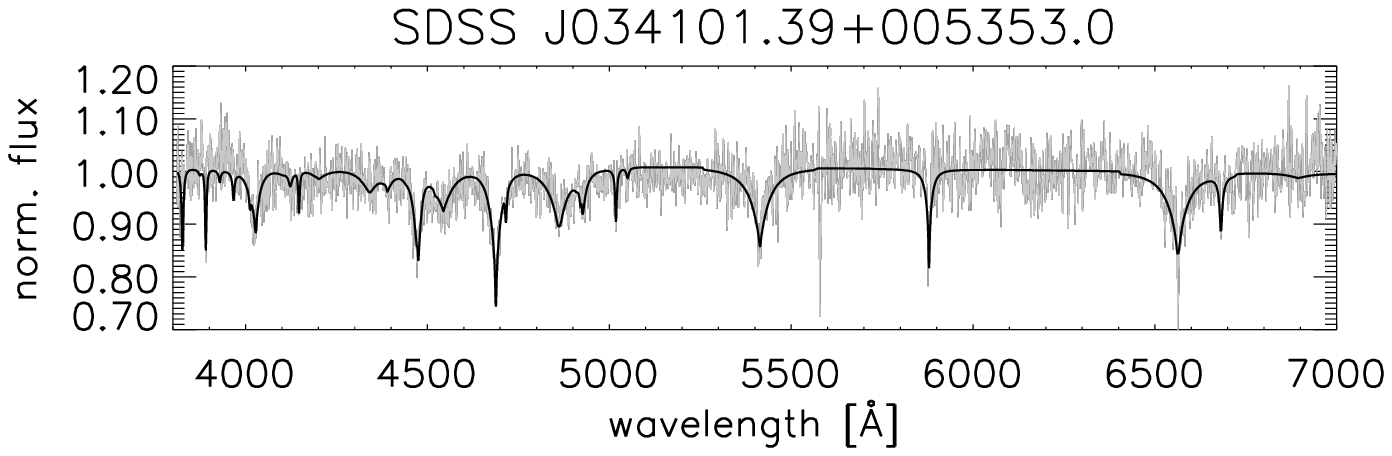}{5cm}{0}{47}{47}{-38}{550}
  \plotfiddle{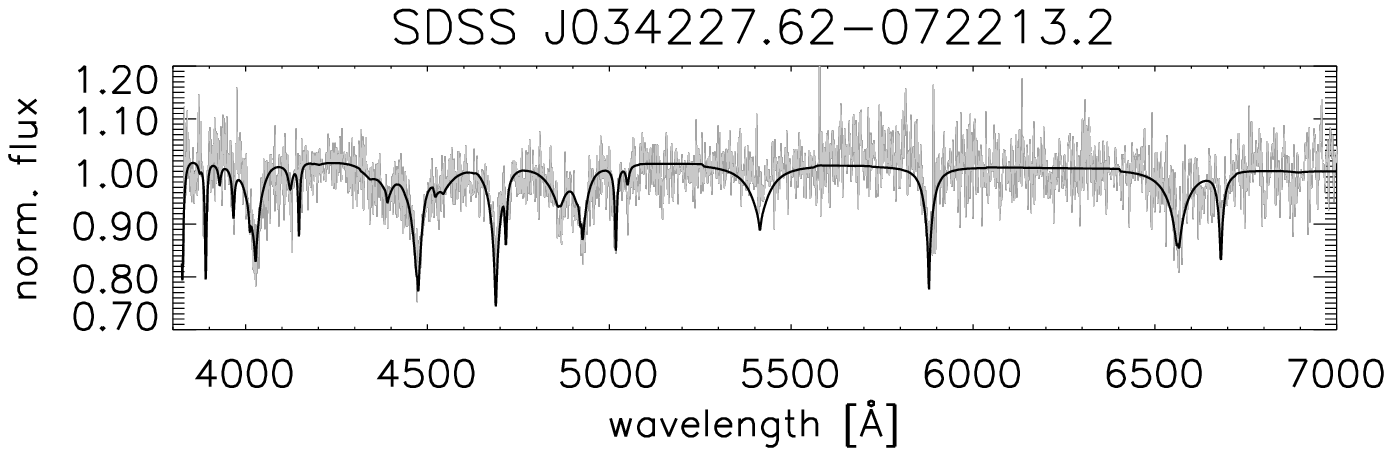}{5cm}{0}{47}{47}{-223}{640}
  \vspace{-11.25in}
  \caption{DO spectra in order of descending temperature (left-right
  top-bottom). The dark lines represent the model spectra.}
  \label{fig:sdh1}
  \plotfiddle{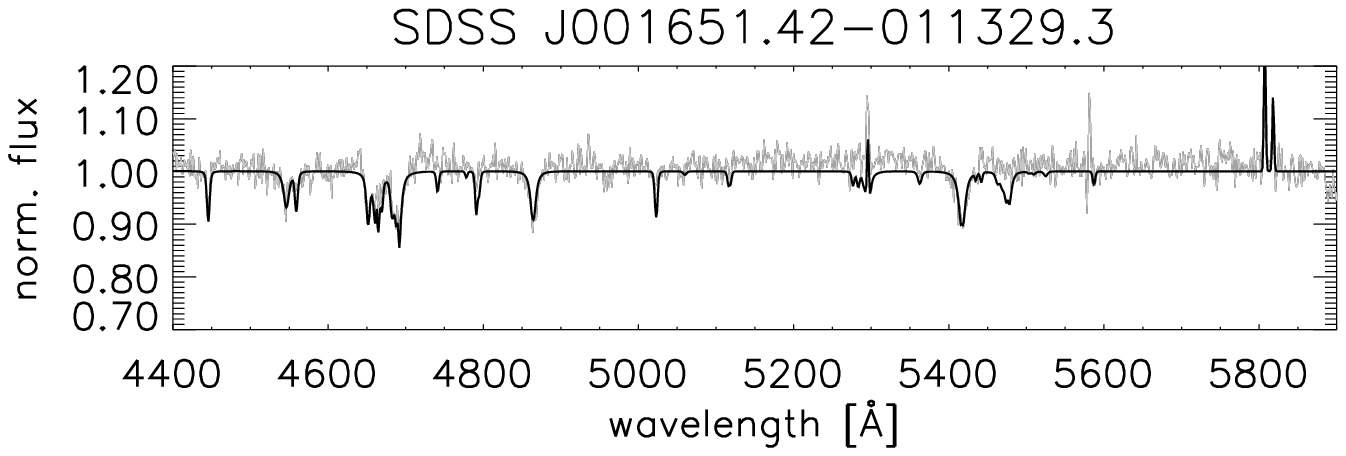}{5cm}{0}{47}{47}{-223}{-105}
  \plotfiddle{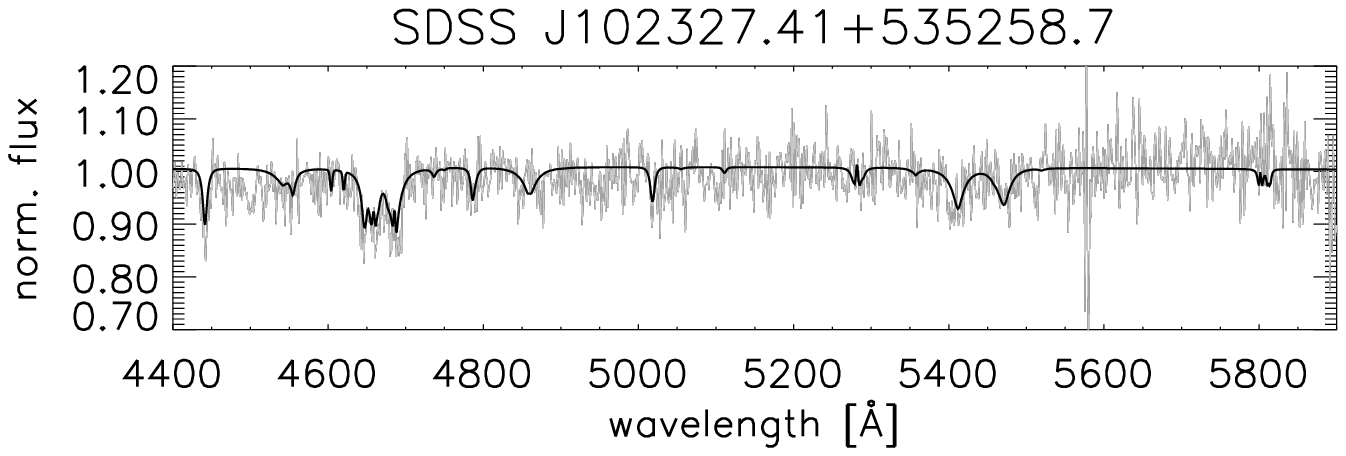}{5cm}{0}{47}{47}{-38}{50}
  \plotfiddle{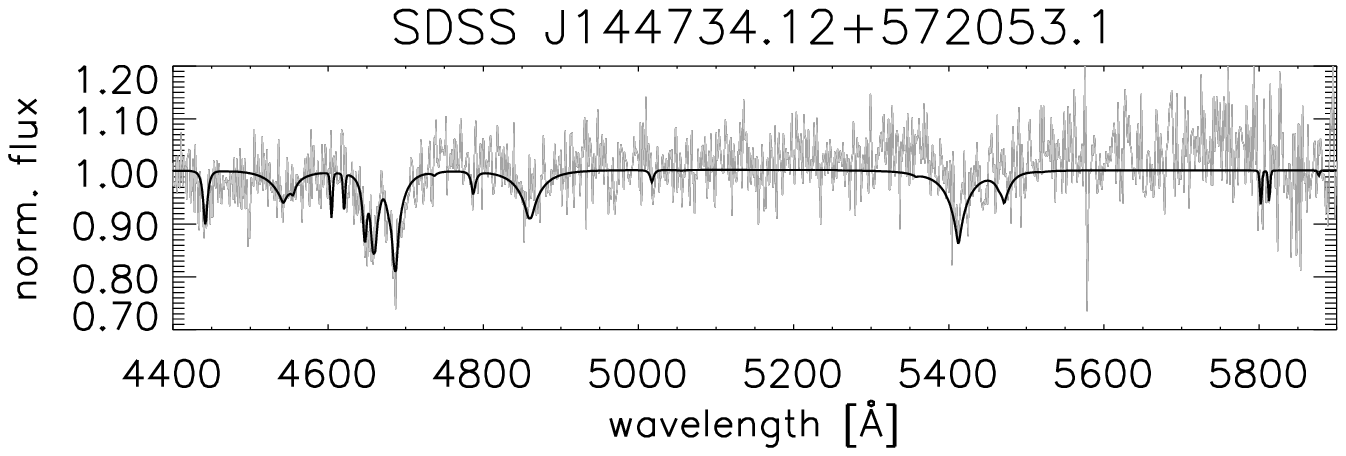}{5cm}{0}{47}{47}{-223}{135}
  \plotfiddle{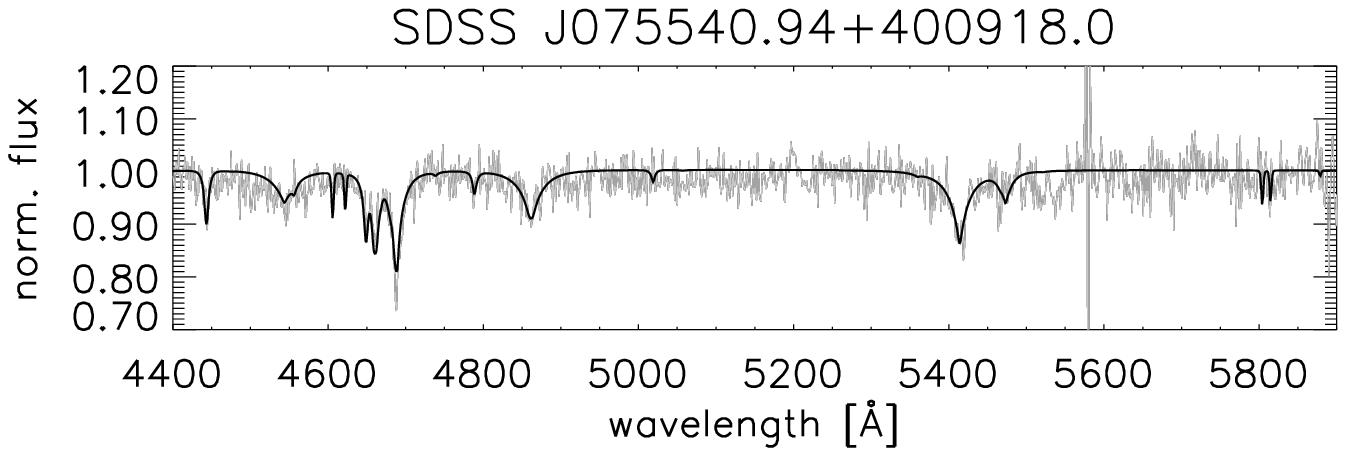}{5cm}{0}{47}{47}{-38}{290}
  \plotfiddle{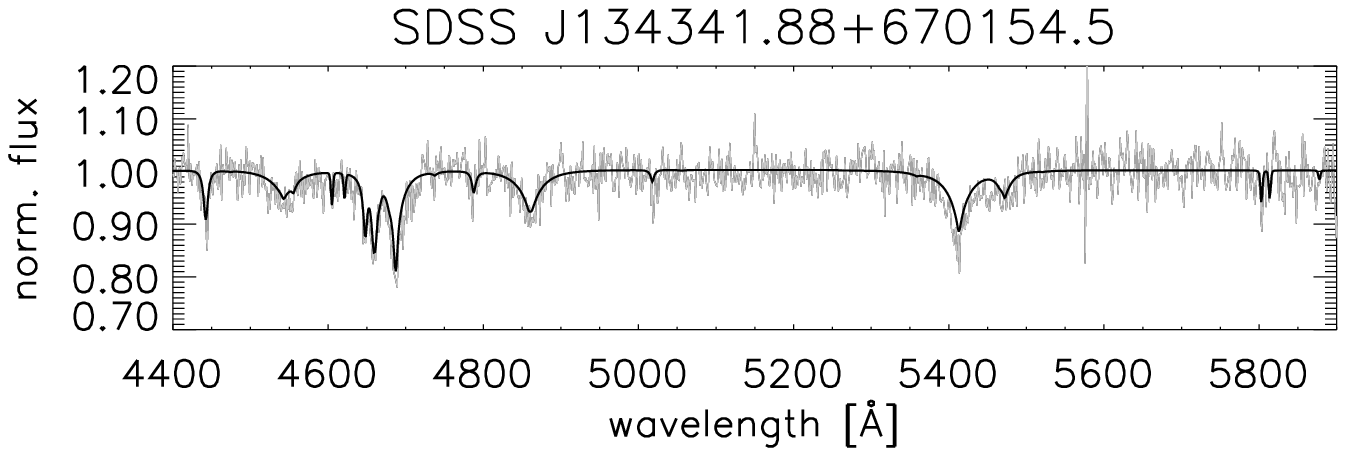}{5cm}{0}{47}{47}{-223}{380}
  \plotfiddle{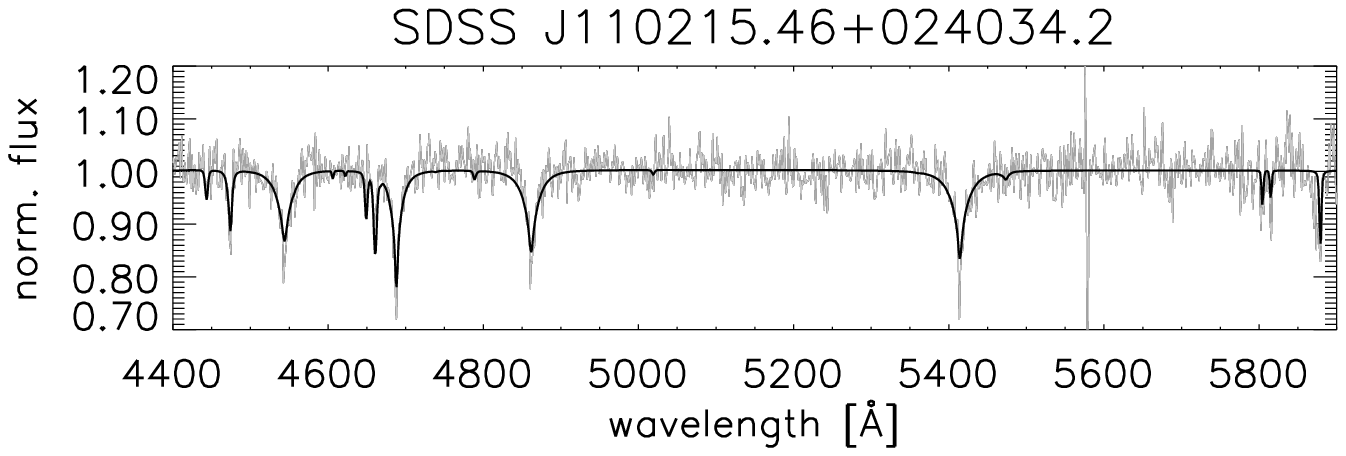}{5cm}{0}{47}{47}{-36.5}{535}
  \vspace{-9.8in}
  \caption{PG1159 spectra in order of descending temperature (left-right
  top-bottom). The dark lines represent the model spectra. SDSS
  J110215.46+024034.3 is a sdO rather than a PG1159 star.}
  \label{fig:sdh2}
\end{figure}

\begin{figure}[H]
  \plotfiddle{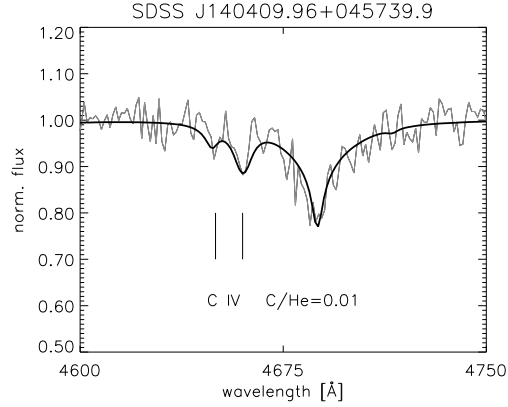}{5cm}{0}{120}{120}{-180}{-440} 
  \vspace{-.2in}
  \caption{Detected C IV lines in the DO white dwarf SDSS
  J140409.96+045739.9. The dark line represents the model spectrum.}
  \label{fig:sdh3}
\end{figure}

In order to analyse the DO stars, we calculated NLTE model
atmospheres with detailed H-He model atoms (Figure
\ref{fig:sdh1}). The model grid ranges
from an effective temperature of $45\,000$ K to $100\,000$ K and a
logarithmic gravity of 7.0 to 8.2. The helium abundance is fixed to
a ratio of He/H = 99. All specified abundance ratios are ratios of
particle numbers. The spectra are taken from SDSS DR1
\citep{2004A&A...417.1093K} and DR2. The models are fitted using a
$\chi^2$ minimization without interpolation. The analyses are
preliminary only. \mbox{SDSS J131724.75+000237.4 = HE1314+0018} has
already been analysed by \citet{2004A&A...424..657W}. For this star we
cannot find a good fit due to unknown reason
\citep[see][]{2004A&A...424..657W}. SDSS J131724.75+000237.4 and
\mbox{SDSS J140409.96+045739.9} provide detectable C IV lines (Figure
\ref{fig:sdh3}).
\newpage

\newpage

For the PG1159 stars we calculated atmospheres using detailed
H-He-C-O model atoms (Figure \ref{fig:sdh2}). The model grid ranges
from  T$_{\rm eff} =
55\,000$ -- $150\,000$ K and $\log g$ [cgs] = 6.4 -- 7.6. The 
abundances are fixed to values He/H = 100 and C/He = 0.01, 0.03, 
0.1, 0.3, 0.6. The fits were done as described above.
\par
One of the objects is an sdO star rather than a PG1159
star (i.e. SDSS J110215.46+024034.2). Most of the PG1159 stars are
cooler ones and moreover most of the stars show moderate C
abundances. O is undetectable in the cooler PG1159 stars. \mbox{SDSS
J001651.42$-$011329.3} shows an abundance of O/He=0.04.

\vspace{-.6in}

\section{Resulting Atmospheric Parameters for DO White Dwarfs and
  PG1159 Stars}
            
From the analyses of the stellar spectra we find the resulting
atmospheric parameters that are shown in the following plots and tables.
            
\begin{figure}[H]
  \plottwo{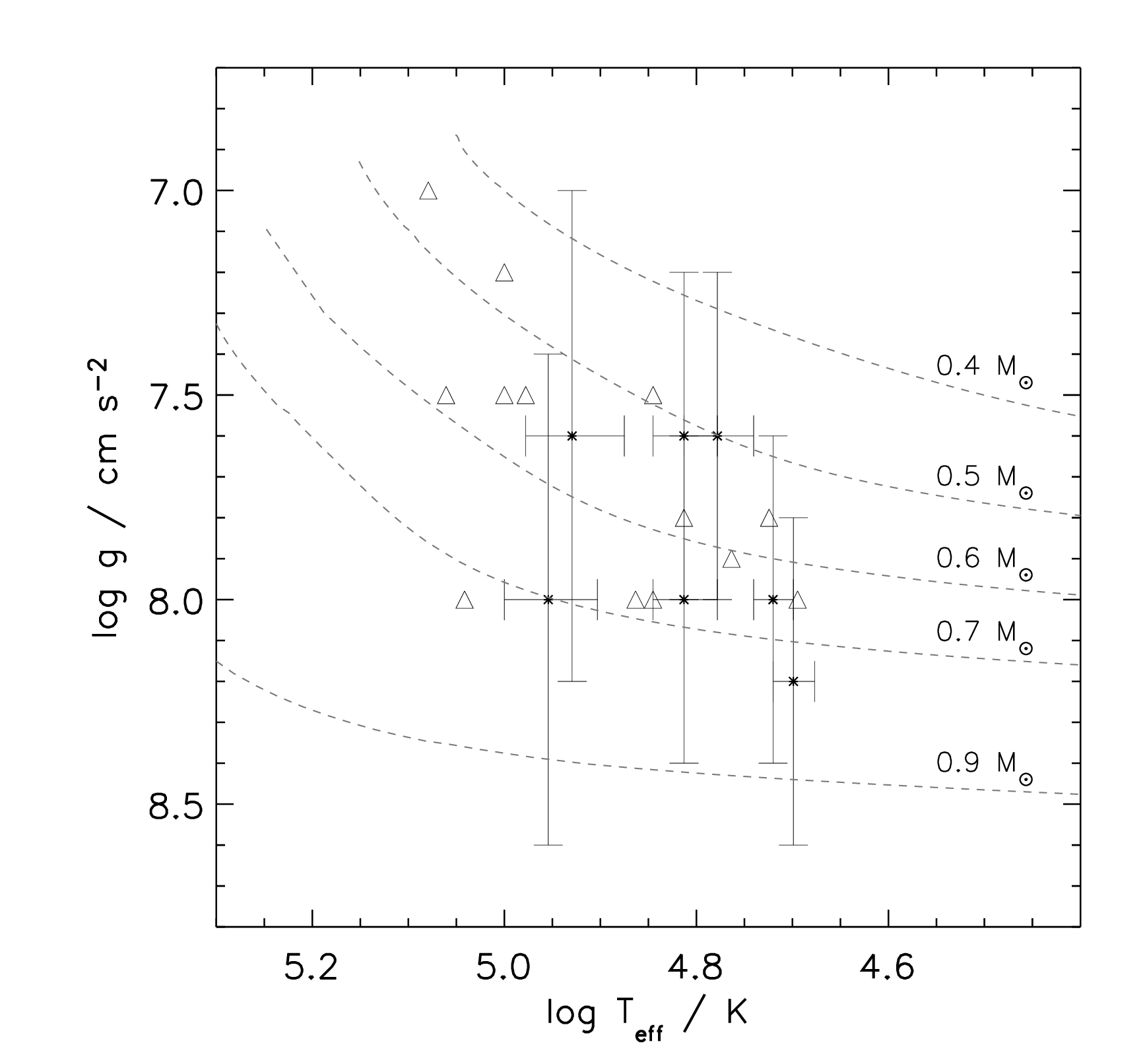}{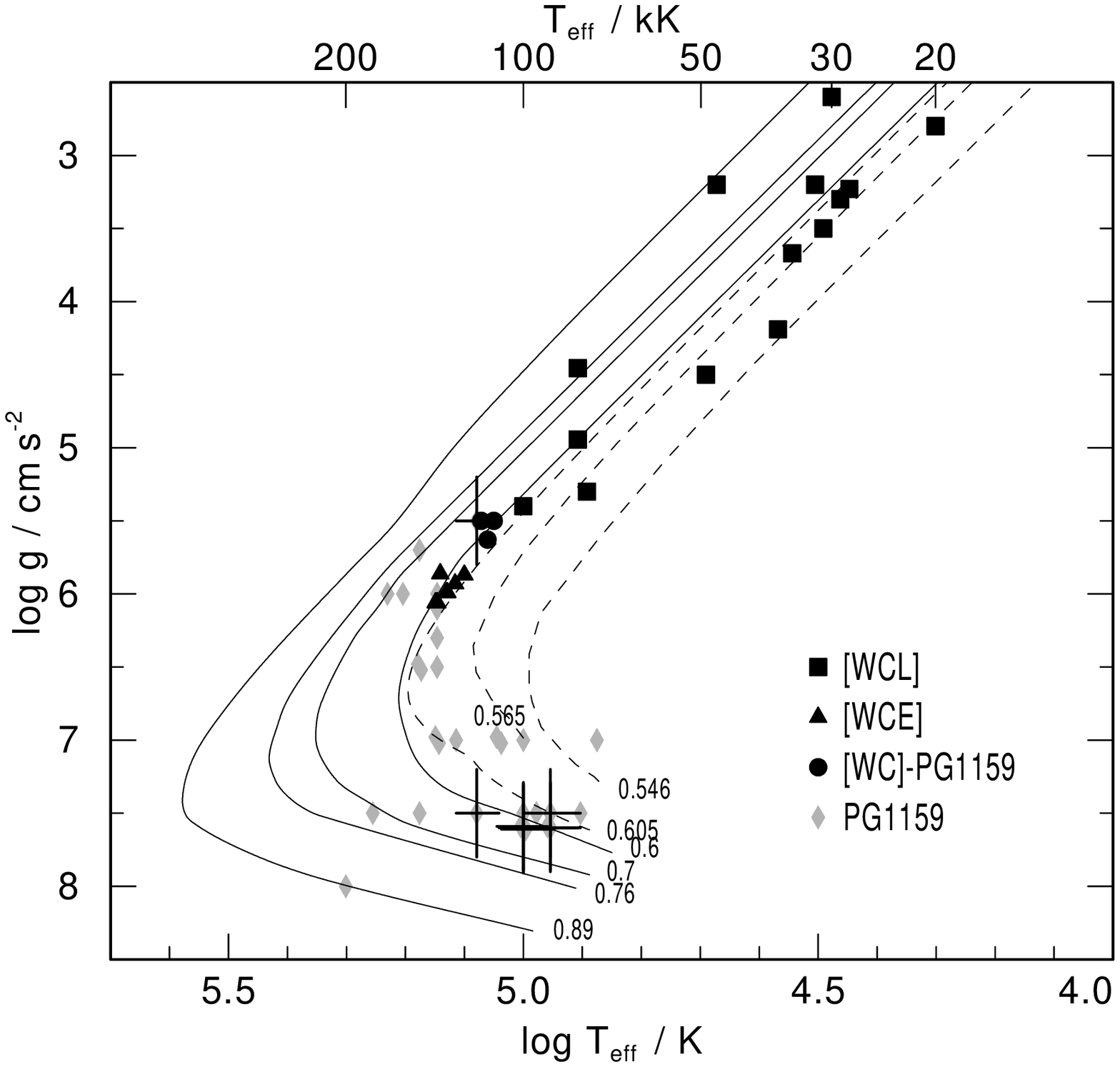}
  \caption{Positions of DO white dwarfs compared with evolutionary
    tracks from \citet{1995LNP...443...41W} and those of the PG1159
    stars with tracks from \citet{1995A&A...299..755B},
    \citet{1983ApJ...272..708S} and \citet*{1986ApJ...307..659W}. The
    triangles in the left plot represent the 14 known DOs
    \citep[see][]{1996A&A...314..217D}.}
\end{figure}

\begin{table}[H]
  \caption{Atmospheric parameters of DOs and PG 1159 stars.}
  \smallskip
  \begin{center}
    {\small
      \begin{tabular}{crcc}
        \noalign{\smallskip}
        \tableline
        \noalign{\smallskip}
        DO white dwarf& T$_{\rm eff}$ & $\log g$ &
        M \\
        &  [K] & [cgs] & [M$_{{\small{\odot}}}$] \\
        \noalign{\smallskip}
        \tableline
        \noalign{\smallskip}
        SDSS J091433.61+581238.1 &  90 000 & 8.0 & 0.70 \\
        SDSS J154752.33+423210.9 &  85 000 & 7.6 & 0.56 \\
        SDSS J140409.96+045739.9 &  65 000 & 8.0 & 0.67\\
        SDSS J155356.81+433228.6 &  65 000 & 7.6 & 0.51\\
        SDSS J131724.75+000237.4 &  60 000 & 7.6 & 0.50\\
        SDSS J034101.39+005353.0 &  52 500 & 8.0 & 0.65\\
        SDSS J034227.62$-$072213.2 &  50 000 & 8.2 & 0.75\\
        \noalign{\smallskip}
        \tableline
        \noalign{\smallskip}
        PG1159 star & T$_{\rm eff}$ & $\log g$ & C/He\\
        &  [K] & [cgs] & [numb. frac.]\\
        \noalign{\smallskip}
        \tableline
        \noalign{\smallskip}
        SDSS J001651.42$-$011329.3 & 120 000 & 5.5 & 0.20 \\
        SDSS J102327.41+535258.7 & 120 000 & 7.0 & 0.30 \\
        SDSS J144734.12+572053.1 & 100 000 & 7.6 & 0.05 \\
        SDSS J075540.94+400918.0 &  100 000 & 7.6 & 0.03 \\
        SDSS J134341.88+670154.5 &  90 000 & 7.6 & 0.05\\
        \noalign{\smallskip}
        \tableline
        \noalign{\smallskip}
        SDSS J110215.46+024034.2 &  55 000 & 6.4 & 0.01\\
        \noalign{\smallskip}
        \tableline
      \end{tabular}
    }
  \end{center}
\end{table}
                         
Apparently there are two types of the PG1159 stars: extremely C-rich
ones (C/He $\approx 0.3$) and less C-rich ones (C/He $\approx
0.03$). The PG1159 star SDSS J001651.42$-$011329.3 is in an unusual
position among progenitor objects (\mbox{[WC]-PG1159} stars).

\vspace{1.in}
     
\acknowledgements{ S. D. H\"ugelmeyer would like to thank the
  organizers of the workshop for financial support. Thanks to T. Stahn
  and J. Steiper for helping to calculate DO model atmospheres.

\end{document}